 \documentclass[preprint]{aastex}

\usepackage{graphicx,palatino}

\shorttitle{Radio-Quiet Quasars towards the Northern HDF}
\shortauthors{Impey and Petry}

\begin{document}


\title{Radio-Quiet Quasars in the Direction \\
    of the Northern Hubble Deep Field}


\author{Chris Impey and Cathy Petry}
\affil{Steward Observatory, The University of Arizona,
    Tucson, AZ 85721}



\begin{abstract}
We match quasars discovered in a multi-color survey centered on the northern
Hubble Deep Field (HDF) with radio sources from an ultra-deep radio survey.
Although 3 out of 12 quasars are detected at a level below
0.2\ mJy at 1.4 GHz, all of the quasars in the search area are radio-quiet 
by the criterion $L_{\rm r} < 10^{25} h^{-1}_{50}$ W Hz$^{-1}$. We combine this 
information with other radio surveys of quasars so as to break the degeneracy 
between redshift and luminosity. In the redshift range $0.02 < z <  3.64$, the
radio-loud fraction increases with increasing optical luminosity, consistent
with some degree of correlation between the non-thermal optical and radio
emissions. More tentatively, for low luminosity quasars in the range $-22.5
< M_B < -25$, the radio-loud fraction decreases with increasing redshift. We 
can infer from this that the radio luminosity function evolves more slowly 
than the optical luminosity function. The mechanism that leads to strong radio
emission in only a small fraction of quasars at any epoch is still unknown.
\end{abstract}




\section{Introduction}

Nearly forty years after their discovery, it is still not clear what
determines the degree of radio emission from quasars. Although quasars
were first discovered at radio wavelengths \citep{sch63}, it has long
been known that most quasars emit a very small fraction of their total
power in the radio \citep{san65}. When radio observations had limited
sensitivity, it made sense to talk about two populations: radio-loud
and radio-quiet. However, \citet{kel89,kel94} detected about 90\% of 
objects in the mostly low redshift Palomar-Green Bright Quasar Survey
(BQS). This survey revealed that (a) the radio luminosity range of the 
BQS is almost disjoint with that of radio-selected quasars, (b) the ratio 
of radio to optical power is continuous and spans at least five orders 
of magnitude, and (c) while most quasars are radio-quiet, few if any 
are truly radio-silent.

In the absence of a physical understanding of the radio emission in most
quasars, it is not clear how best to define radio-loudness. Many workers
have used the ratio of radio-to-optical power, $R$, to characterize the 
degree of radio emission \citep{kel89,hoo96,bis97}. This has an advantage
as a distant-independent quantity, but its interpretation is only physically
meaningful if radio and optical luminosity are correlated. \citet{pea86}
and \citet{sto92} showed that this is unlikely to be the case. Another
common definition of radio-loud is based on a radio luminosity boundary,
$L_{\rm r} > 10^{25} h^{-1}_{50}$ W Hz$^{-1}$ \citep{hoo96,bis97}. Radio
luminosity can be interpreted in terms of a physical model. The choice of 
log $L_{\rm r} = 25$ for radio-loud is convenient since it corresponds to 
a typical $\sim 1$ mJy detection limit for a quasar with $M_B = -26$ at 
$z \sim 1$. Radio-loudness is an arbitrary concept in the sense that no 
one has convincingly demonstrated that the distributions of either radio
luminosity or radio-to-optical power are bimodal.

Demographic studies of the radio emission from quasars are affected by 
the fact that the upper bounds in a detection experiment are correlated 
with redshift. Different samples must therefore be combined to break the
degeneracy between luminosity and redshift. The only way to study low
luminosity quasars at high redshift is to combine a faint optical survey
with extremely sensitive radio observations. We have chosen to do this in 
the direction of the northern Hubble Deep Field (HDF), a small region of sky 
which is at the locus of a wide range of extragalactic studies \citep{wil96}
. In \S2 of this paper, we use a recently published deep radio
survey to measure the radio luminosities of quasars found in a one square
degree region centered on the HDF. In \S3, we combine this sample
with other quasar surveys to study the radio-loud fraction as a function 
of redshift and luminosity.  We present conclusions in \S4.  Modelling of
quasar evolution and speculation as to the nature of the radio emission
mechanism is deferred to a future paper. We use $H_0 = 50$
kms$^{-1}$ Mpc$^{-1}$ and $q_0 = 0.5$ ($\Lambda=0$) throughout for the 
calculation of luminosity.

\section{Identifying Optically Faint Quasars with Weak Radio Sources}

The northern Hubble Deep Field, along with its southern twin, has 
motivated a broad array of projects designed to measure the luminous 
content of the universe out to high redshift. We have been carrying out
multi-color surveys of optically faint quasars in one square degree regions
centered on each of the deep fields. The quasars will be used to thread 
these volumes with absorption sightlines, and to compare the location of
the absorbers with the large scale structures measured by pencil beam
redshift surveys of galaxies \citep{coh96,ste96,hog98}. Our first paper
reported 30 quasars in the magnitude range $17.6 < B < 21.0$ and the
redshift range $0.44 < z < 2.98$ \citep{liu99}. An independent group is 
working on quasar selection in a smaller region centered on the HDF
\citep{van00}, and they find an additional 4 quasars with $B\lesssim21$. 

To complement the optical imaging with the Hubble Space Telescope (HST),
\citet{ric98} have carried out a deep radio survey of the HDF down to a 
5$\sigma$
limit of 9 $\mu$Jy at 8.5 GHz. More recently, \citet{ric00} presented a
catalog of 372 radio sources (1.4 GHz) within 20 arcmin of the HDF, 
with a 3$\sigma$ sensitivity limit ranging from $40\ \mu$Jy at the center 
of the field to 230\ $\mu$Jy at a radius of 20 arcmin.
The strongest sources in the
sky are radio galaxies and quasars in roughly equal numbers. At flux 
levels a million times lower, the radio source population is likely to
be composed primarily of star-forming disk galaxies, with a relatively
small component of quasars and other AGN \citep{ric98}. 
The advantage of such a sensitive
radio survey for our purposes is the fact that quasars at redshifts of
$1 < z < 2$ can be detected down to the traditional boundary between
radio-loud and radio-quiet quasars.

Figure 1 shows the distribution of HDF radio sources from \citet{ric00}, 
plotted as open circles, along with the distribution of spectroscopically
confirmed quasars, plotted as stars. There are 34 quasars within a one
square degree area centered on the HDF, 30 of which come from \citet{liu99}
and 4 of which are added from \citet{van00}. Since both studies used a UVX
selection technique and found almost all their quasars in the magnitude
interval $19 < B < 21$, it is appropriate to combine them. 

In a detection experiment such as this, we must bear in
mind the fact that the HDF is not a truly random patch
of sky in terms of numbers of strong radio sources \citep{wil96}.
To allow very deep radio observations, the field
was chosen such that there was no source stronger than
1 mJy in the primary beam of the VLA at 8.5 GHz. The
full sample of \citet{ric98} has 29 sources
within 4.6 arcmin of the HDF center.
However, our experiment is carried out over
the substantially larger field of the 1.4 GHz observations
\citep{ric00}.
For the observed median spectral index of 0.63 \citep{ric00},
1~mJy at 8.5 GHz corresponds to 3.1 mJy at 1.4 GHz.
There are 9 sources stronger than 3.1 mJy
over the 1.4 GHz field of view, or a surface density of
0.0072 sources per square arcminute. The expected number
of sources in the smaller region of the 8.5 GHz data is therefore
only 0.48. Thus the pre-ordained absence of a
source in the central region is not a large perturbation
on the statistics of our experiment.  In addition, the central
region is only 5.3\% of the total area of our identification
experiment.

A potentially 
more serious issue is incompleteness in the quasar surveys. The radio
survey is uniform down to a particular flux limit, but multicolor quasar
selection is not perfectly efficient, and not all candidates brighter than 
$B = 21$ have had confirming spectroscopy. This means that the region could
contain radio sources identified with unrecognized quasars, which would 
bias the identification process to lower radio-to-optical flux ratios. 
In practice, we believe this is not an important effect. 
The UVX technique is known to be efficient out to $z\sim 2.5$, which
encompasses the quasars used in this study \citep{hew94}.
The observed surface density of 34 deg$^{-2}$ is in excellent agreement with 
the highest surface densities ever measured down to this magnitude level 
\citep{zit92,hal96}.

There are 12 quasars in the region sampled by the deep VLA observations,
as outlined by the circle in Figure 1. These quasars can be identified by
positional coincidence with the radio sources. The rms astrometric errors 
on the radio and optical positions are 0.03$\arcsec$ and 0.5$\arcsec$,
respectively \citep{ric00,van00}. With accurate positions and a low surface
density of quasars and radio sources, there is a small probability of a
chance coincidence within three times the rms positional error. However,
Figure 1 shows that the HDF sources are centrally concentrated, 
due to attenuation by the primary beam pattern. 
A statistical test confirms that the radial distribution of radio sources
departs from the expectation for a random distribution at more than the
99.9\% confidence level. By contrast, the quasar distribution within one
square degree is consistent with a random distribution. This means that the
probability of a chance coincidence is higher towards the center of the HDF, 
and it indicates that we should carry out a more rigorous procedure before
declaring a radio counterpart.

The results of the identification procedure are summarized in Table 1. For 
each of the 12 confirmed quasars within the circular region measured by the 
VLA, we select the nearest radio source as a potential counterpart. The first
five columns of Table 1 give the name, J2000 position, $B$ magnitude and
redshift of the 12 quasars. The simulation procedure takes into account the
different surface densities of quasars and radio sources and the fact that 
the radio source distribution shows a radial gradient. We randomly reposition
the 34 quasars within the one square degree and repeat this 1500 times. 
The radial distance to the field center, $r$,  and the angular distance to the 
nearest
neighbor radio source, $\theta_{\rm nn}$, are saved for each of these 51,000 
random quasar placements. Then, for each of the 12 quasars, we form the 
distribution of $\theta_{\rm nn}$ for random quasars falling in an annulus 100 
arcseconds wide centered on the radial distance of each quasar from the center 
of the HDF. 

The nearest neighbor distributions for this random experiment are shown in
Figure 2. Each panel in the figure is labelled with four items. The first 
is the quasar name. The second is the radial distance of that quasar from 
the center of the HDF. The histograms show how the mean nearest neighbor
distance depends on $r$ --- the surface density of radio sources is highest
near the HDF, so the nearest neighbor distances will on average be smallest.
The overall number of trials was chosen so that there were at least 1000
measurements in each histogram. 
The third item labelled is the actual offset of that quasar from the nearest 
radio source, $\theta_{\rm nn}$ (column 8 in Table 1), marked by a vertical 
dotted line in the figure. 
The fourth item is the fraction of trials in the random placement experiment
where the quasar had an angular distance of $\theta_{\rm nn}$ or smaller 
from the nearest radio source. Equivalently, it is the probability, $p$, 
that the actual nearest neighbor is a chance counterpart. This number can 
easily be converted into a percentage confidence limit on the identification,
($1-p$)/100.

Three out of 12 of the quasars have weak radio sources within 1.5 arcseconds,
or within three times the sum of the radio and optical position errors added
in quadrature. These quasars are therefore identified with greater than 
99.5\% confidence. It is clear that the identification procedure is robust,
since none of the other nine quasars have radio sources within 15 arcseconds.
The probability $p$ is strongly bimodal; three quasars have $p < 0.002$ 
(taking into account the statistical accuracy due to the number of points
generated for each Monte Carlo histogram), and
the other nine have $p > 0.18$. To ensure uniqueness, we inspected the 2nd
and 3rd nearest neighbors for the three quasars with close companions.  The 
nearest neighbor is by far the most likely true counterpart
with $p>0.23$ in all cases, except for the 2nd nearest neighbor to J123800+6213
which has $p=0.02$. 
Radio fluxes for the three new identifications are listed in the last column of
Table 1. 
The other nine quasars are assigned a 3$\sigma$ upper bound 
based on the quoted formal completeness limit of $40\ \mu$Jy 
with a scaling by the instrumental gain as presented in Figure 3 of
\citet{ric00}.

\citet{ric98} found no quasar or AGN identifications with sources from their
8.5 GHz survey.  We can use our results to put an approximate upper bound on
the quasar component of the 1.4 GHz survey.  Only 3 out of 372, or about 1\%,
faint radio sources are identified with quasars.  We can scale this number up
by a factor of 6 to account for 50\% incompleteness due to color selection,
and at most a factor of 3 increase in optical quasar counts down to $B\sim24$
\citep{boy88}.  Thus, we predict no more than $\sim5\%$ of ultrafaint radio
sources can be quasars of {\em any} optical brightness.  This conclusion is
firm because the optical survey is deep enough to have seen all the way
through the quasar luminosity function for all reasonable cosmologies, and our
existing identifications and limits are near the lower bound of the
distribution of radio-to-optical flux ratios.

\section{Radio-Loud Fraction as a Function of Redshift and Luminosity}

As stated in the introduction, we prefer to characterize quasars by their
radio luminosity rather than their radio-to-optical luminosity ratio, since
emissions in the two wavelength regimes appear to not be closely related.
Early workers suspected that there were distinct radio-loud and radio-quiet
categories of quasars \citep{kel89,mil90,sto92,vis92}. However, the onset of 
the radio-quiet population happens to correspond to the typical flux limit 
of a VLA snapshot, so the evidence for bimodality was never compelling. As
sample sizes have increased, the evidence for two populations has weakened
\citep{hoo96,gol99,bis97}. \citet{kel89} established the procedure that has
become standard for most papers in this field, defining a monochromatic 
radio luminosity in the rest frame and assuming that the radio emission 
is isotropic, with units of W Hz$^{-1}$. Radio luminosity is a continuous
variable in the quasar population, but $L_{\rm r} > 10^{25} h^{-1}_{50}$ 
W Hz$^{-1}$ is a useful definition of radio-loudness, because it corresponds 
to the beginning of the tail of the distribution. It is also roughly equivalent
to the 
definition of \citet{gol99}, $L_{\rm r} > 10^{24} h^{-1}_{50}$ W Hz$^{-1}$
sr$^{-1}$. 

In this section, we combine the HDF quasar sample with other samples to 
explore the dependence of the radio-loud fraction on redshift and luminosity.
The calculation of all distance-dependent quantities assumes $H_0 = 50$
km s$^{-1}$ Mpc$^{-1}$ and $q_0 = 0.5$ ($\Lambda = 0$). Rest frame radio
luminosities are calculated for an observed flux at 5 GHz. This matches
the frequency of BQS data; radio k-corrections for other samples assume
a radio spectral index $\alpha = -0.5$ ($S_\nu \propto \nu^\alpha$). Rest
frame absolute magnitudes for the BQS are adopted from \citet{kel89}. The
calculation of rest frame absolute magnitude for all other samples uses the
k-corrections calculated by \citet{hoo95} using a composite LBQS spectrum.
The typical uncertainty in these derived quantities for a particular quasar 
are $\sim 0.6$ mag for $M_B$ \citep{hoo95} and $\sim 0.15$ for log $L_{\rm r}$
\citep{all92}. The large uncertainty in optical luminosity is due to source
variability and the large dispersion in rest frame UV to optical spectral
indices. We note that these uncertainties are small compared with the overall
range of radio ($\times 10^5$) and optical ($\times 10^3$) luminosity.

Figure 3 shows the coverage of the various samples in terms of linear look-back time and optical luminosity. Equivalent values of redshift are also
plotted. The diamonds show the 114 quasars of the optically bright BQS
\citep{kel89,kel94}. The open circles show the 367 quasars with radio
measurements from the LBQS \citep{vis92,hoo95,hoo96}. We choose not to 
augment this with additional matches in large area surveys like FIRST
\citep{whi97} and NVSS \citep{con98}, because those surveys have 2-6 times 
lower sensitivity limits. The stars show 87 quasars from the Edinburgh survey
\citep{gol99}, which covers a similar region of the redshift-luminosity plane 
to the LBQS. The twelve new HDF quasars are shown as filled circles in Figure 3.
The dotted lines show the plane divided up into equal sized regions of
look-back time and absolute magnitude.  The highest redshift slice encompasses
the epoch of maximum quasar space density.  The lower redshift slices span the
time over which the space density has declined by two orders of magnitude to
the present epoch.  The radio-loud fractions in these intervals of look-back
time and optical luminosity are investigated  in the subsequent analysis.

The intercomparison of samples with bright and faint optical limits allows 
us to break the degeneracy between redshift and luminosity which exists for 
any single, flux-limited sample. With a bright magnitude limit and wide area
sky coverage, the BQS includes some of the most optically luminous quasars
known. However, evidence that the BQS is incomplete \citep{gol92,gra00}, along
with the possibility that the incompleteness depends on radio properties
\citep{pea86,mil93,hoo96}, leads us to be cautious in drawing conclusions
that depend only on the BQS. The LBQS is a homogeneous survey with a well
understood selection function \citep{hew95}. Compared to the BQS, the LBQS
reaches lower optical luminosities at a given redshift and higher redshifts
at a given optical luminosity. Even though the HDF sample is small, Figure
3 shows that it covers an important range of parameter space at low optical
luminosity and moderate redshift.

Figure 4 shows absolute magnitude plotted against radio luminosity for the
three samples that cover approximately the same redshift range:  the HDF, LBQS
and Edinburgh samples. Symbols are the same as in Figure 3. Upper limits are
plotted as arrows for the HDF quasars; a box outlines the region containing
the upper limits for LBQS and Edinburgh quasars.  Two conclusions can be drawn
from this plot.  First, the HDF quasars are extremely weak radio emitters.
All twelve are below the traditional threshold for a radio-loud quasar,
$L_{\rm r}=10^{25}$ W Hz$^{-1}$.  In fact, all but one of the twelve 
are a factor of ten weaker
than this limit, and 2 out of 12 are about a factor of 100 weaker than this
limit.  It is clear from Figure 4 that the great sensitivity of the HDF radio
survey pushes the parameter space of radio luminosity for moderate redshift
quasars.  Also, there is no strong correlation between radio and optical
luminosity for these samples. 

We have identified six regions in Figure 3 
for investigating the demographics of the radio-loud fraction. The sequence
from 1 to 2 to 3 represents the cosmic evolution of the radio-loud fraction,
in equal bins of one third of the look-back time, in the low luminosity range
$-25 < M_B < -22.5$. The sequence from 1 to 4 to 6 represents the change in
the radio-loud fraction with increasing optical luminosity by a factor of a
thousand, in the redshift range $0.02 < z < 3.64$.  We include the BQS sample
in this comparison, noting that these quasars only contribute to the bin at
low redshift and low luminosity (region 3) and the bin at high redshift and
high luminosity (region 6).

Table 2 shows the statistics from these four samples combined, for three
different definitions of the radio-loud fraction: log $L_{\rm r} > 26$, log
$L_{\rm r} > 25$, and log $L_{\rm r} > 24\ h^{-1}_{50}$ W Hz$^{-1}$. All of 
the radio surveys have upper limits, but log $L_{\rm r} > 26$ is above the
limits in any of the surveys we are using. The criterion log $L_{\rm r} > 25$
matches the most common definition in the literature, and it is immune from 
the nature of the radio non-detections for quasars less luminous than $M_B =
-26.5$ (see dashed box that defines upper limits in Figure 4). We note that 
all but one of the HDF quasars are radio-quiet at the level of 
log $L_{\rm r} = 24$, but the nature of the upper limits becomes important 
above a luminosity of $M_B = -24$. The footnotes in Table 2 give the number
of quasars for which the designation radio-loud or radio-quiet is uncertain
because the detection upper limit lies above the selected boundary in 
$L_{\rm r}$ .  A lower 
limit on the radio-loud fraction can be estimated by assuming all of the 
uncertain quasars are radio-quiet --- these values are also included in the 
footnotes to Table 2.

The data in the table show no trend for evolution in the radio-loud fraction 
of low luminosity quasars with cosmic time. A low radio-loud fraction of 
0.05-0.10 is seen at all epochs (for $L_{\rm r}>25$). In agreement with 
\citet{gol99}, we see an increase in the
radio-loud fraction with increasing optical luminosity. In the redshift 
interval $0.84 < z < 3.64$, the radio-loud fraction rises by a factor of
3-4 over a span of a factor of a thousand in optical luminosity. The low
luminosity HDF quasars provide the strongest tether on this result, because
1/9 are radio-loud at any of the three radio thresholds. 

\section{Conclusions}
We have identified quasars selected from a multi-color survey with
radio sources from a very deep catalog of a region centered on the 
northern Hubble Deep Field. The identification procedure accounts for
the non-uniform radial distribution of radio sources over this region.
Although limited by small number statistics, the result of the comparison
is new information on the radio properties of low luminosity quasars at
moderate redshift. We reach the following conclusions:

(1) Twelve optically selected quasars with $B < 21$ and $0.4 < z < 2.6$
are in the region within 20 arcmin of the HDF, which was covered in the
deep radio survey of Richards (1999). Three of the quasars are identified
with weak radio sources at greater than a 99.9\% confidence level. The 
remaining nine are below a $3\sigma$ limit of 40 to 230 $\mu$Jy at 1.4 GHz.

(2) The HDF quasars are extremely weak radio-emitters. All but one of the
twelve have radio luminosities more than an order of magnitude below the 
traditional boundary between radio-loud and radio-quiet quasars, 
$L_{\rm r} = 10^{25}$ W Hz$^{-1}$.
Two of the twelve are about two orders of magnitude
below that boundary, and are the lowest radio luminosity quasars ever measured
at these redshifts.  

(3) There is no strong evidence that radio and optical luminosity are 
correlated,
a conclusion that is essentially independent of the nature of the radio
non-detections. 
If we make a plausible assumption about the completeness of the optical quasar
selection, no more than $\sim5$\% of the radio source population above 
$230\ \mu$Jy can be quasars at any optical brightness level.

(4) The fraction of quasars more radio luminous than 10$^{26}$ W Hz$^{-1}$
is 3-5\% (or 5-10\% above $10^{25}$ W Hz$^{-1}$), independent of redshift over
the range $0.02 < z < 3.64$. The fraction
of quasars above $10^{24}$ W Hz$^{-1}$ in radio luminosity shows slight evidence for
an increase with increasing redshift. While the quasar space density evolves
over two orders of magnitude, the probability of a quasar being a
strong radio source is independent of epoch.

(5) The radio-loud fraction of quasars shows a marginally significant increase
by a factor of three, going from low to high optical luminosity. 
This modest effect, combined with the lack of any strong
correlation between radio and optical luminosity, means that the quasar
radio luminosity function cannot be much flatter than the optical luminosity
function.

\acknowledgments
We acknowledge useful discussions with Pippa Goldschmidt, Eric Hooper, Charles
Liu, and Eric Richards.  We are grateful to the referee for pointing out a 
significant error in the first version of this paper.  CDI is grateful to 
the NSF for support under grant AST-9803072.

\clearpage



\begin{center}
\includegraphics[scale=0.6,angle=-90]{fig1.ps}
\figcaption[fig1.ps]{The one square degree field centered on the northern  
Hubble Deep Field (J2000).  The 34 quasars are represented by filled stars
\citep{liu99,van00}. The 40\arcmin\ field of the radio survey \citep{ric00} is 
represented by the large circle, and the 372 radio sources are plotted as 
small open circles.}
\label{fig1}
\end{center}

\clearpage
\begin{center}
\includegraphics[scale=0.56]{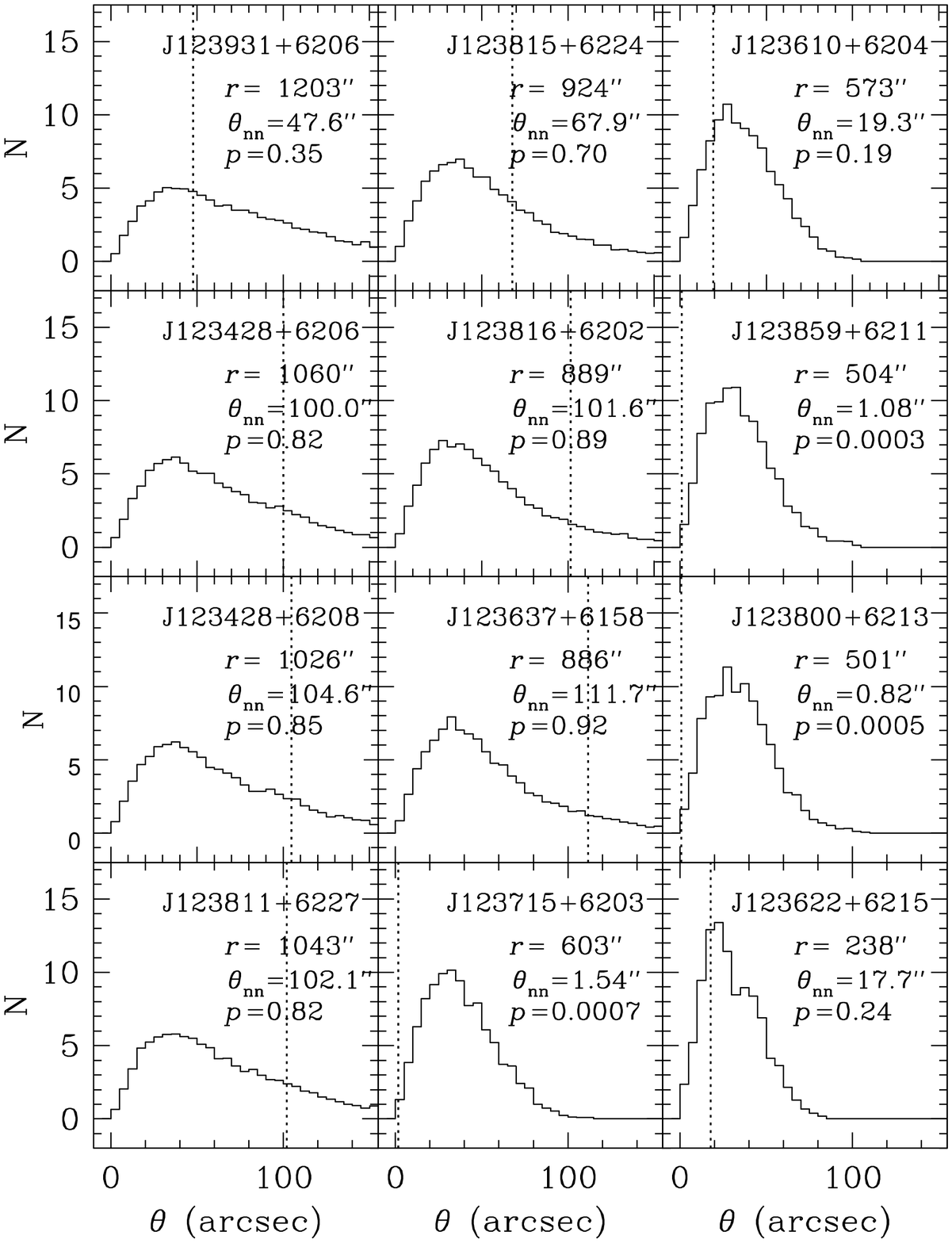}
\figcaption[fig2.ps]{
The results of a Monte Carlo simulation to test the significance of quasar
identifications.  The histograms show the distribution of distance to the
nearest radio source for random placements of the quasar (measured at the
appropriate radial distance from the HDF).
Each panel is labelled with the name of the quasar, the radial distance of the 
quasar from the center of the field in arcseconds, the angular distance to the
nearest radio source, $\theta_{\rm nn}$ (denoted graphically by the dotted 
vertical line), and the probability that the nearest neighbor is a chance 
counterpart (equal to the fraction of the histogram with $\theta < \theta_{\rm
nn}$).
}
\label{fig2}
\end{center}

\clearpage
\begin{center}
\includegraphics[scale=0.75,angle=-90]{fig3.ps}
\figcaption[fig3.ps]{The coverage of the four samples in the optical
luminosity--look-back time plane.  The diamonds are the 114 BQS quasars, the 
small open circles are the 367 LBQS quasars, the stars are the 87 quasars from 
the Edinburgh survey, and the 12 HDF quasars are shown as large filled circles.
The radio-loud fraction is investigated for the numbered regions defined by 
the dotted lines.
}
\label{fig3}
\end{center}

\clearpage
\begin{center}
\includegraphics[scale=0.75,angle=-90]{fig4.ps}
\figcaption[fig4.ps]{The optical luminosity -- radio luminosity plane.  The 
small open circles are the LBQS quasars, the stars are the Edinburgh survey 
quasars, and the HDF quasars are shown as large filled circles.  The arrows
are upper limits for HDF quasars.  The region that encompasses the upper
limits for the LBQS and Edinburgh surveys is shown as a dashed box.
}
\label{fig4}
\end{center}

\clearpage

\begin{deluxetable}{ccccrcrrr}
\footnotesize
\tablecaption{Quasars Within the HDF-N Radio Survey Area\label{tab1}}
\tablewidth{0pt}
\tablehead{
\colhead{Name} & \colhead{$\alpha$}   & \colhead{$\delta$} &
\colhead{$B$} &
\colhead{$z$} &
\colhead{Cross ID} &
\colhead{$r$\tablenotemark{a}} &
\colhead{$\theta_{nn}$\tablenotemark{b}} &
\colhead{$S_{1.4}$\tablenotemark{c}} \\
\colhead{} & \colhead{(J2000)}   & \colhead{(J2000)} &
\colhead{} &
\colhead{} &
\colhead{} &
\colhead{(\arcsec)} &
\colhead{(\arcsec)} &
\colhead{($\mu$Jy)} 
}
\startdata
J123428+6208 & 12 34 28.24 & 62 08 23.8  &  21.03  &  1.355\tablenotemark{*} & \nodata & 1026 & 105  & $<149$ \\
J123428+6206 & 12 34 28.41 & 62 06 32.1  &  20.47  &  1.793\tablenotemark{*} & \nodata & 1060 &  100 & $<162$ \\
J123610+6204 & 12 36 10.24 & 62 04 35.3  &  19.86  &  1.74~~  & 12  & 573 & 19.3 & $<62$ \\
J123622+6215 & 12 36 22.89 & 62 15 27.4  &  20.50  &  2.58~~  & 13  & 238 & 17.7 & $<42$ \\  
J123637+6158 & 12 36 37.45 & 61 58 15.6  &  18.95  &  2.52~~  & 14   & 886 & 112 & $<107$ \\
J123715+6203 & 12 37 15.96 & 62 03 24.5  &  20.18  &  2.05~~  & 15  & 603 & 1.55& 109  \\
J123800+6213 & 12 38 00.85 & 62 13 36.8  &  19.16  &  0.44~~  & 17  & 501 & 0.82 & 190 \\
J123811+6227 & 12 38 11.99 & 62 27 27.5  &  20.76  &  0.77~~  & 18  & 1043 & 102 & $<155$ \\
J123815+6224 & 12 38 15.46 & 62 24 40.7  &  21.33  &  1.75~~  & 19  & 924  & 67.9 & $<117$ \\
J123816+6202 & 12 38 16.06 & 62 02 09.2  &  19.18  &  1.00~~  & 20   & 889  & 102 & $<108$ \\
J123859+6211 & 12 37 59.51 & 62 11 03.4  &  18.77  &  0.910\tablenotemark{*} & \nodata & 504 & 1.08 & 85 \\
J123931+6206 & 12 39 31.44 & 62 06 20.1  &  19.54  &  1.19~~  & 25  & 1203  & 47.6 & $<230$ \\
\enddata

\tablenotetext{a}{The radial distance in arcseconds of each quasar from the 
center of the HDF.}
\tablenotetext{b}{The angular distance in arcseconds from each quasar to the 
nearest radio source.}
\tablenotetext{c}{The radio flux at 1.4 GHz \citep{ric00} of the source
identified with each quasar. The upper bounds represent $3\sigma$ limits,
calculated using the gain curve of \citet{ric00}.}

\tablecomments{Quasar names, positions, and $B$ magnitudes are from
\citet{van00}.  The redshift is quoted from \citet{liu99}, except
where marked with an asterisk to denote values taken from \citet{van00}.  The 
sixth column is the cross-identification number from Table 1 of \citet{liu99}.}

\end{deluxetable}

\clearpage

\begin{deluxetable}{ccrrr}
\footnotesize
\tablecaption{Radio-Loud Fraction\label{tab2}}
\tablewidth{0pt}
\tablehead{
\colhead{Region} & 
\colhead{Range} & 
\colhead{Fraction with} &
\colhead{Fraction with} &
\colhead{Fraction with} \\ 
\colhead{in Figure 3} & 
\colhead{} & 
\colhead{$\log L_{\rm r} > 26$} &
\colhead{$\log L_{\rm r} > 25$} &
\colhead{$\log L_{\rm r} > 24$} \\ 
}
\startdata
\tableline
\multicolumn{5}{c}{Redshift} \\
\tableline
1 & $3.64<z<0.84$     & $0.048\pm0.049$  & $0.095\pm0.070$ & $0.300\pm0.197$\tablenotemark{a} \\
2 & $0.84<z<0.27$     & $0.025\pm0.012$  & $0.080\pm0.023$ & $0.143\pm0.034$\tablenotemark{b} \\
3 & $0.27<z<0.0$      & $0.033\pm0.024$  & $0.049\pm0.029$ & $0.066\pm0.034$ \\
\tableline
\multicolumn{5}{c}{Optical Luminosity} \\
\tableline
1 & $-22.5<M_b<-25.0$ & $0.048\pm0.049$  & $0.095\pm0.070$ & $0.300\pm0.197$\tablenotemark{a} \\
4 & $-25.0<M_b<-27.5$ & $0.077\pm0.021$  & $0.118\pm0.028$\tablenotemark{c} & $0.674\pm0.162$\tablenotemark{d} \\
6 & $-27.5<M_b<-30.0$ & $0.164\pm0.056$  & $0.353\pm0.097$\tablenotemark{e} & $0.897\pm0.242$\tablenotemark{f} \\
\enddata

\tablenotetext{a}{11/21 cases are uncertain,  $f_{\rm lim} \gtrsim 0.143\pm0.088$. 
$\ ^{\rm b}$~22/162 cases are uncertain, $f_{\rm lim} \gtrsim 0.123\pm0.029$.  
$\ ^{\rm c}$~12/181 cases are uncertain, $f_{\rm lim} \gtrsim 0.110\pm0.026$.  
$\ ^{\rm d}$~138/181 cases are uncertain, $f_{\rm lim} \gtrsim 0.160\pm0.032$.
$\ ^{\rm e}$~10/61 cases are uncertain,  $f_{\rm lim} \gtrsim 0.295\pm0.079$.  
$\ ^{\rm f}$~32/61 cases are uncertain,  $f_{\rm lim} \gtrsim 0.426\pm0.100$.}

\tablecomments{The footnoted values indicate that for the corresponding region, 
the designation of radio-loud or radio-quiet is uncertain for some number of 
quasars because the limiting radio flux is higher than the bound 
$\log L_{\rm r}>24,25,\rm or\ 26$.  An approximate lower limit on the 
radio-loud fraction, $f_{\rm lim}$, is computed assuming these quasars have
$L<L_{\rm r}$ and are radio-quiet.
}
\end{deluxetable}

\end{document}